\documentclass{aastex63}  

\usepackage{graphicx}
\usepackage{epstopdf}
\usepackage{subfigure}
\usepackage{booktabs}
\usepackage{url}
\usepackage{amssymb}
\makeatletter

\newcommand{\Rmnum}[1]{\expandafter\@slowromancap\romannumeral #1@}
\makeatother

\shorttitle{$\gamma$-ray QPO in PKS 0521-36}
\shortauthors{Zhang et al.}

\begin{document}

\title{A Quasi-periodic Oscillation in the $\gamma$-ray Emission from the Non-blazar Active Galactic Nucleus PKS 0521-36}

\correspondingauthor{Dahai Yan}
\email{yandahai@ynao.ac.cn}
\correspondingauthor{Li Zhang}
\email{lizhang@ynu.edu.cn}

\author{Haiyun Zhang}
\affiliation{Department of Astronomy, Key Laboratory of Astroparticle Physics of Yunnan Province, Yunnan University, \\Kunming 650091, China}

\author{Dahai Yan}
\affiliation{Key Laboratory for the Structure and Evolution of Celestial Objects, Yunnan Observatory, Chinese Academy of Sciences,\\
	Kunming 650011, China}

\author{Pengfei Zhang}
\affiliation{Department of Astronomy, Key Laboratory of Astroparticle Physics of Yunnan Province, Yunnan University, \\Kunming 650091, China}

\author{Shenbang Yang}
\affiliation{Department of Astronomy, Key Laboratory of Astroparticle Physics of Yunnan Province, Yunnan University, \\Kunming 650091, China}

\author{Li Zhang}
\affiliation{Department of Astronomy, Key Laboratory of Astroparticle Physics of Yunnan Province, Yunnan University, \\Kunming 650091, China}

\begin{abstract}

Long-term $\gamma$-ray variability of a non-blazar Active Galactic Nucleus (AGN) PKS 0521-36 is investigated by using {\it Fermi}-LAT pass 8 data covering from 2008 August to 2021 March.
The results show that the histogram of the $\gamma$-ray fluxes follows a log-normal distribution.
Interestingly, in the analysis of $\sim$5.8-year (from MJD 56317 to 58447) LAT data between two outbursts (occurring during 2012 October and 2019 May respectively),
a quasi-periodic oscillation (QPO) with a period of $\sim1.1$ years ($\sim5\sigma$ of significance) is found in the Lomb-Scargle Periodogram (LSP), 
the Weighted Wavelet Z-transform (WWZ) and the REDFIT results.
This quasi-periodic signal also appears in the results of Gaussian process modeling the light curve.
Therefore, the robustness of the QPO is examined by four different methods.
This is the first $\gamma$-ray QPO found in a mildly beamed jet.
Our results imply that the $\gamma$-ray outbursts play an important role in the formation of the $\gamma$-ray QPO.
\end{abstract}


\section{Introduction} \label{sec:intro}

It is generally accepted that Active Galactic Nuclei (AGNs) existing in the center of a small fraction of galaxies are powerful emitters for the entire galaxies \citep[e.g.,][]{2017A&ARv..25....2P}. 
Usually, an AGN is powered by a central super massive black hole (SMBH) and emits variable emission covering from radio to high-energy gamma-ray bands \citep[e.g.,][]{2010ApJ...724.1517B,2016ARA&A..54..725M}. Non-thermal GeV-TeV $\gamma$-rays from AGNs are produced in their powerful relativistic jets. 

The largest class of AGNs detected by {\it Fermi} Large Area Telescope ({\it Fermi}-LAT) is of blazar type \citep{2020ApJ...892..105A}. A typical blazar has relativistic jet oriented at very small angle ($\theta$ $\textless 5^{\circ}$) to the line of our sight \citep{2015ApJ...810...14A}, and therefore its jet emission is strongly Doppler-boosted \citep{1995PASP..107..803U}. Rapid flares on timescale of a few minutes in blazars have been observed at GeV $\gamma$-ray energies \citep[e.g.,][]{2016ApJ...824L..20A,2019ApJ...877...39M} and TeV energies \citep[e.g.,][]{2011ApJ...730L...8A}. This kind of rapid flare suggests an extremely compact $\gamma$-ray emission region which likely locates in the broad-line region (BLR) of flat spectrum radio quasars (FSRQs). However, no signs of $\gamma$-ray absorption by BLR photons in $\gamma$-ray spectrum suggests that $\gamma$-ray emission region should be outside of the BLR \citep[e.g.,][]{2018MNRAS.477.4749C}. To reconcile the observations of fast variability and simultaneous energy spectrum, new models are proposed \citep[see][for recent reviews]{2019Galax...7...20B,2019ARA&A..57..467B}. Moreover, timing analysis, such as the exploration of periodicity, would provide important information about the nature of these sources. Quasi-periodic variabilities of blazars in $\gamma$-ray energies have been investigated, and possible quasi-periodic variabilities in more than 20 blazars have been reported based on analyses of {\it Fermi}-LAT data \cite[e.g.,][]{2014ApJ...793L...1S,2016AJ....151...54S,2017A&A...600A.132S,2015ApJ...810...14A,2017MNRAS.471.3036P,2017ApJ...835..260Z,2017ApJ...845...82Z,2017ApJ...842...10Z,2018NatCo...9.4599Z,2019MNRAS.487.3990B,2020ApJ...896..134P,2020ApJ...891..163Z}.
However, the cause for the $\gamma$-ray quasi-periodic variabilities remains controversial. \citet{2018ApJ...867...53Y} investigated the relation between $\gamma$-ray variability amplitude and spectral index for PG 1553+113, and suggested that the $\gamma$-ray quasi-periodic variability of PG 1553+113 \citep{2015ApJ...810...14A} could be caused by a periodic modulation in Doppler factor. Several possible explanations have been proposed for the $\gamma$-ray quasi-periodic variabilities in blazars, for instance, presence of a binary system of SMBHs \citep{2016ApJ...819L..37V, 2018ApJ...854...11T}, modulation in the accretion flow feeding the jet \citep{2003MNRAS.344..468G}, and a periodic lighthouse effect \citep{1992A&A...255...59C}.

Seventy non-blazar AGNs are included in the Fourth Catalog of AGNs detected by {\it Fermi}-LAT \citep[4LAC;][]{2020ApJ...892..105A}. Among them, there are 41 radio galaxies (RGs), 9 narrow-line seyfert 1 galaxies (NLSy1s), 2 steep-spectrum radio quasars (SSRQs), 5 compact steep spectrum radio sources (CSSs), 1 Seyfert galaxy, and 11 other AGNs. 6 NLSy1s, 7 RGs, 3 CSSs, 1 SSRQ, and 3 other AGNs are found to be variable \citep{2020ApJ...892..105A}. The jet properties of NLSy1 are similar to those of FSRQs, but with lower powers \citep[e.g.,][]{2013MNRAS.436..191D,2015MNRAS.446.2456D,2016MNRAS.463.4469D,2013ApJ...774L...5Z,2019ApJ...872..169P}. $\gamma$-ray flares on a daily timescale are found in several NLSy1s \citep{2019Galax...7...87D}.  RGs are misaligned AGNs, and therefore the jet Doppler boosting effect is not strong. Intraday variability is found in the {\it Fermi}-LAT data of the RG NGC 1275 \citep{2018A&A...614A...6S,2018ApJ...860...74T}. All of the non-blazar AGNs are not bright enough to have good sampling in their long-term $\gamma$-ray light curves, except for NGC 1275 and PKS 0521-36. The $\gamma$-ray variability and spectral evolution of NGC 1275 have been studied in detail \citep[e.g.,][]{2017ApJ...848..111B,2018A&A...614A...6S,2018ApJ...860...74T}. Here, we focus on the $\gamma$-ray properties of PKS 0521-36. A combination of $\gamma$-ray properties of blazars and non-blazar AGNs can provide insightful information on AGN jet physics.

PKS 0521-36 is a peculiar non-blazar AGN with broad emission lines in the optical and ultraviolet bands and the steep radio spectrum \citep{1995A&A...303..730S,1986ApJ...302..296K}.
\cite{2015MNRAS.450.3975D} raised a possibility that PKS 0521-36 is an intermediate source between broad-line radio galaxy and SSRQ. 
It was previously classified as an N galaxy, then a BL Lac object \citep{2015MNRAS.450.3975D}.
The latest LAT source catalog (4FGL; \citealt{2020ApJS..247...33A}) suggests that the existing data cannot determine a clear classification of this source. 
Very Long Baseline Array (VLBA) observations reveal that the radio structure of its jet is similar to misaligned AGN \citep{2015MNRAS.450.3975D}. 
Recently, \citet{2019A&A...627A.148A} confirmed that the jet of PKS 0521-36 is not highly beamed.

The jet structure of PKS 0521-36 has been studied widely at frequencies from optical to X-rays \citep[e.g.,][]{1999ApJ...526..643S,1994Msngr..77...49F,1986ApJ...302..296K}.
Because of its prominent radio, optical, and X-ray jet \citep{1979MNRAS.188..415D,1991ApJ...369L..55M,2002MNRAS.335..142B,2017ques.workE..16L},
PKS 0521-36 is considered to be one of the most remarkable extragalactic objects. At $\gamma$-ray energies, 
a strong flare on variability timescale of 12-hour was detected by {\it Fermi}-LAT from PKS 0521-36 in 2010 June \citep{2015MNRAS.450.3975D}. 
Recently, a $\gamma$-ray flare on timescale of $\sim6$-hour in 2012 October was reported by \citet{2019A&A...627A.148A}.

A study on the long-term variability behavior of PKS 0521-36 in $\gamma$-ray energies is lack. 
The characteristics of $\gamma$-ray variability are helpful to understand the nature of this source. 
In this work, we perform a full time-domain analysis on the latest {\it Fermi}-LAT data of PKS 0521-36. 
The paper is organized as follows. The basic introduction of {\it Fermi}-LAT and results of searching for periodicity by different methods are presented in Section~\ref{sec:style}. 
Our discussion of the results is given in Section~\ref{sec:discussion}. Finally, a summary is presented in Section~\ref{sec:summary}.

\section{Data Analysis and Results} \label{sec:style}

\subsection{{\it Fermi}-LAT Likelihood Analysis}
\label{subsec:Fermi}

{\it Fermi}-LAT is a pair conversion $\gamma$-ray detector with wide field of view ($\sim 2.4\rm\ sr$), large effective area ($>8000\ \rm cm^{2}$ at $\sim$1 GeV),
and broad-energy covering from $\sim$ 20 MeV to $>$ 300 GeV. Thanks to its all-sky monitoring capabilities, one can get long-coverage and uniform observations of $\gamma$-ray sources. A detailed description of LAT can be found in \cite{2009ApJ...697.1071A}.

The data used here were collected during $\sim$12.6 years of {\it Fermi}-LAT operation from 2008 August 4 to 2021 March 29 [Modified Julian Day (MJD):
$54682-59302$] in the 0.1-500 GeV energy range. The analysis of data follows the standard criteria for the point-source analysis\footnote{\url{http://fermi.gsfc.nasa.gov/ssc/data/analysis/documentation/Pass8_usage.html}}. We employ the Science Tools package of version v11r05p3 available from {\it Fermi} Science Support Center\footnote{\url{https://fermi.gsfc.nasa.gov/ssc/data/analysis/software}} (FSSC), and use the Pass 8 LAT database in which we choose a region of interest (ROI) of $15^{\circ}$ centered at PKS 0521-36. Two background templates, Galactic and extragalactic diffuse emissions, are modeled with the file gll\_iem\_v07.fits and iso\_P8R3\_SOURCE\_ V2\_v1.txt, respectively. The instrumental response function ``P8R3\_SOURCE\_V2" is adopted in data processing. In order to get events with high probability of being photons, we only keep the SOURCE class events through setting ``evclass = 128, evtype = 3" in the {\it Fermi} tool {\it gtselect}. 
In addition, the events with zenith angle $>90^{\circ}$ are filtered out to avoid contaminating from Earth's limb.
Good time intervals (GTIs) are selected by using the criterion of ``($\rm DATA QUAL>0$)\&\&($\rm LAT CONFIG==1$)" in the {\it Fermi} tool {\it gtmktime}. 
Above settings ensure that we adopt high-quality observations.

\subsection{$\gamma$-ray SED and Light Curves} \label{subsec:tables}

\begin{figure}
\centering
	{\includegraphics[width=11cm]{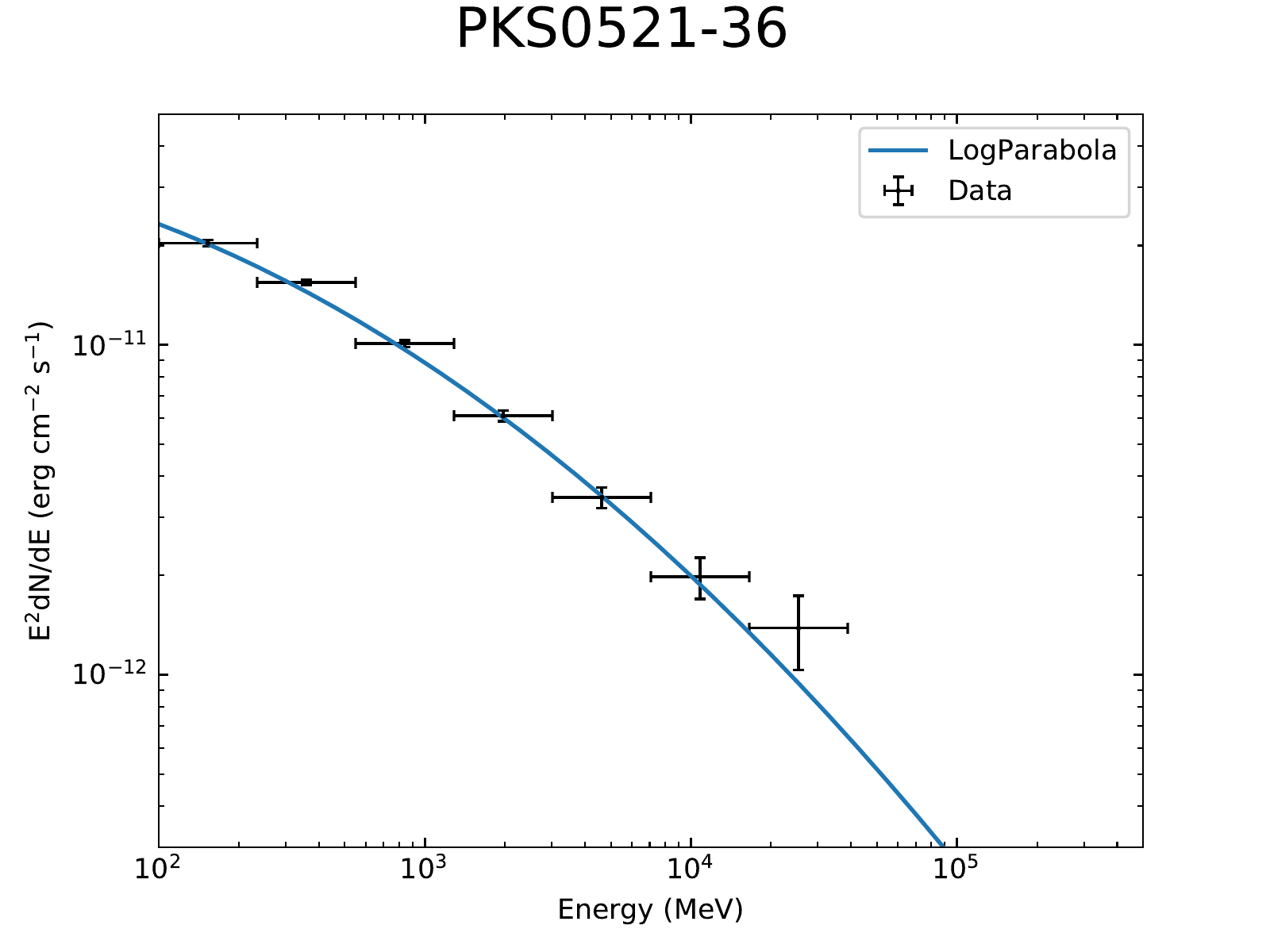}}
	\caption{Average SED of PKS 0521-36 from MJD 54682 to 59302.
	The best-fitting LP model is plotted as blue solid line, and the best-fitting parameters are given in the text.}
	\label{fig:sed}
\end{figure}

In the process of analyzing the entire data, an initial binned maximum likelihood analysis \citep{1996ApJ...461..396M} is performed.
The spectral parameters and the normalization factors of all sources lying within $5^{\circ}$ of PKS 0521-36 as well as the normalization factors of the sources located beyond $5^{\circ}$ but within $10^{\circ}$ from our target are set to be free,
while the sources from ROI but beyond $10^{\circ}$ have their parameters fixed to the values in {\it Fermi}-LAT fourth source catalog \citep{2020ApJS..247...33A}.
In the iterative fittings, the best-fitting results are obtained, and are saved to be a new model file which is the basis of constructing light curves and spectrum energy distribution (SED). The spectral model of LogParabola (LP) form is used here, which is
\begin{equation}\label{eq1}
\frac{dN}{dE}=N_{0}\left(\frac{E}{E_{\rm b}}\right)^{-[\alpha + \beta \log(E / E_{\rm b})]}\ .
\end{equation}
The global fitting results in an integrated average flux of $(1.06\pm 0.02) \times 10^{-7}$\ photons\ cm$^{-2}$\ s$^{-1}$ and a total Test Statistics (TS) value of 18447.2. 
The TS is defined as 2log($L/L_{0}$), where $L$ is the maximum likelihood of the model with a point source at the target position, and $L_{0}$ is the maximum likelihood without the source \citep{1996ApJ...461..396M}.
The best-fitting spectral parameters are the spectral slope $\alpha = 2.43\pm0.02$ and the curvature parameter $\beta=0.05\pm 0.01$.
The average SED of PKS 0521-36 in the whole time period is shown in Figure \ref{fig:sed}.

\begin{figure}
	\centering
	{\includegraphics[width=11cm]{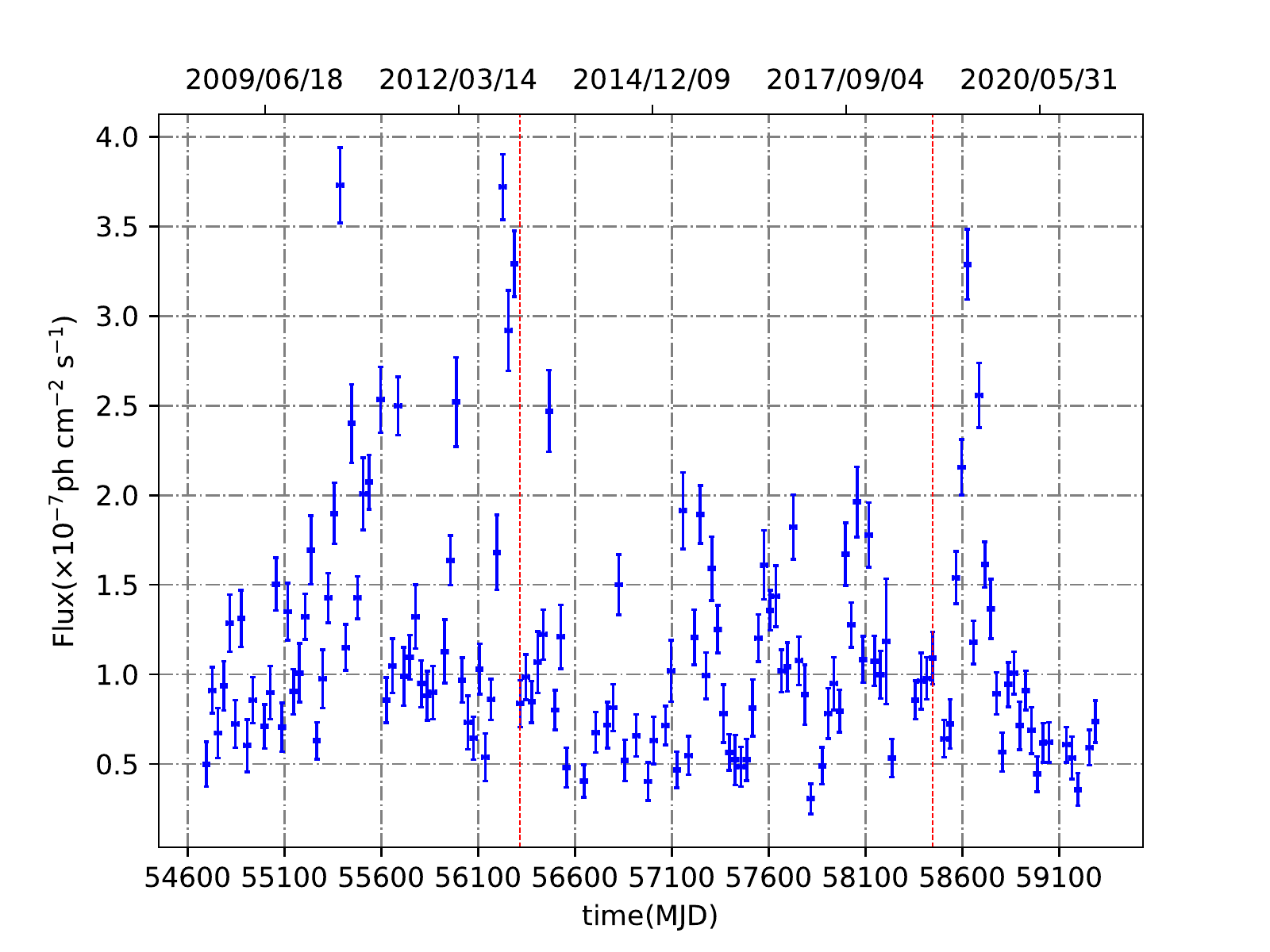}} 
	\caption{LAT light curve of PKS 0521-36 from MJD 54697 to 59287 in the energy range of 0.1 GeV - 500 GeV with 30 days time bin. 
	             Two vertical dashed lines mark the time range of MJD 56317$-$58447, during which periodic modulation is analyzed (see Section \ref{sec:periodicity}).
	} 
	\label{fig:lightcurve-alphafitall}
\end{figure}

The unbinned maximum likelihood optimization is used to produce $\gamma$-ray light curve of PKS 0521-36. 
In each time bin, the new model file mentioned above is employed, but the spectral parameters of all sources are freezed. 
The time bins having $TS<$25 and the number of the model-predicted photon $N_{\rm pred}<$10 are excluded .
This ensures that  the robust detections of the source are adopted in the analysis \citep{2020ApJS..247...27K,2013MNRAS.436.1530R}.
A 30-day binning light curve is produced. The light curve shown in Figure \ref{fig:lightcurve-alphafitall} reveals flux variability over the whole observational period.
Three high flaring activities appear in the light curve. The first is in 2010 June and, has been reported by \cite{2015MNRAS.450.3975D} who also reported the contemporaneous increases of activities observed in optical, UV and X-ray bands. The second flaring occurred in 2012 October has been studied by \citet{2019A&A...627A.148A}, and the latest flaring activity occurred in 2019 May.

\begin{figure}
	\centering
    {\includegraphics[width=11cm]{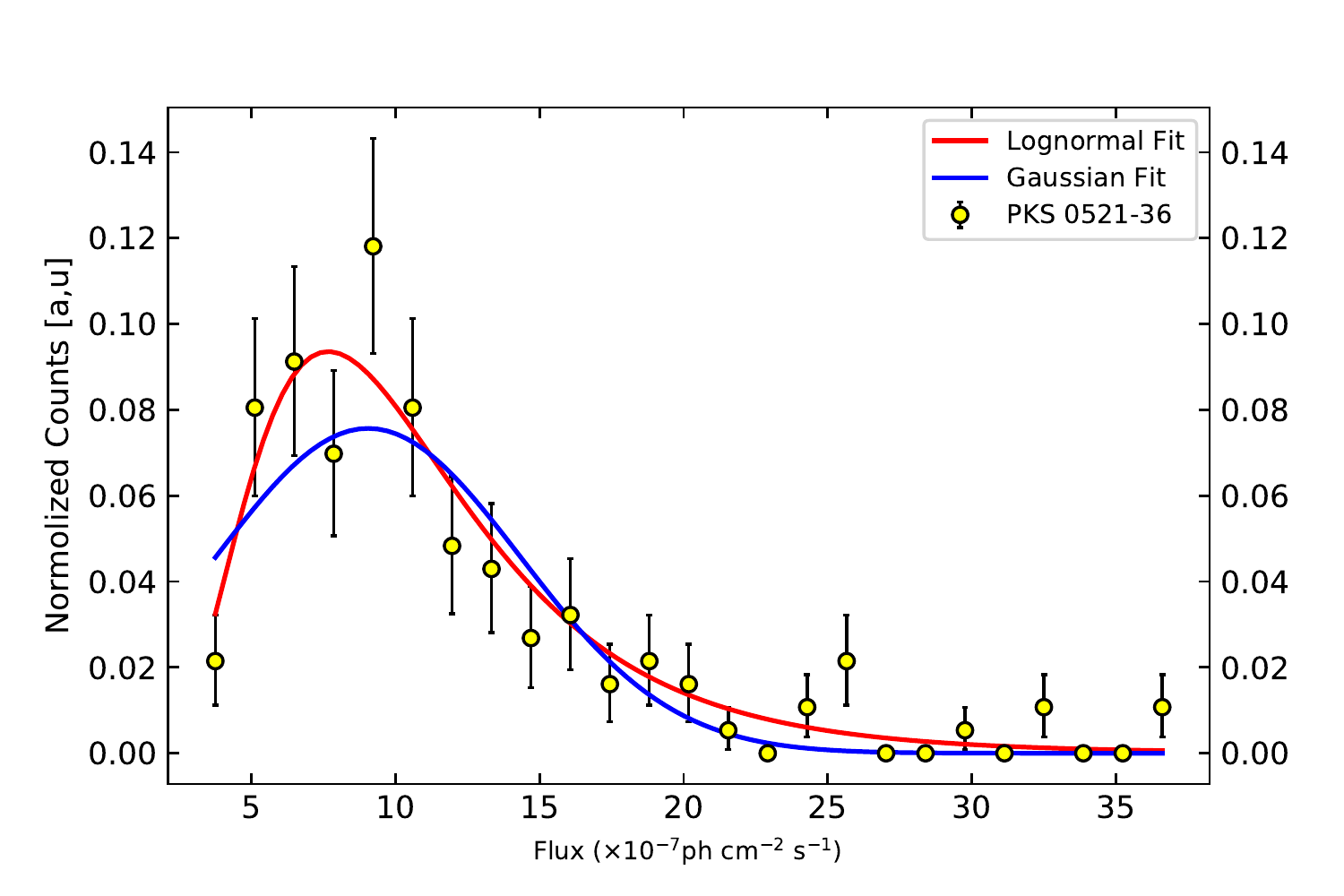}}
    \caption{Normalized histogram of $\gamma$-ray photon fluxes fitted by the log-normal (red solid line) and Gaussian (blue solid line) function.} 
	\label{fig:fluxdistribution}
\end{figure}

\subsection{Flux Distribution } \label{subsec:distribution}

One important property of a time series is its probability density function (PDF). 
One can explore the mechanism that drives variability by analyzing the flux distribution of a light curve \citep[e.g.,][]{2019Galax...7...28R}. 
PDF can be estimated by fitting a histogram of observed data. The distribution of the $\gamma$-ray photon fluxes of PKS 0521-36 is presented in a histogram with 25 bins, 
and is fitted by log-normal and Gaussian functions in the form of
\begin{equation}\label{eq2} L(\phi\mid\mu,\sigma)=\frac{1}{\sqrt{2\pi}\sigma\phi}\exp\left[-\frac{(\log(\phi)-\mu)^{2}}{2\sigma^{2}}\right]\;,
\end{equation}
\begin{equation}\label{eq3}
G(\phi\mid\mu,\sigma)=\frac{1}{\sqrt{2\pi}\sigma}\exp\left[-\frac{(\phi-\mu)^{2}}{2\sigma^{2}}\right]\;,
\end{equation}
respectively. The parameter $\sigma$ is the standard deviation and the parameter $\mu$ is the mean of the distribution. 
The flux histogram and best-fitting results are shown in Figure~\ref{fig:fluxdistribution}. 
The errors of data points are estimated by the method given by \cite{1986ApJ...303..336G}. 
The best-fitting parameters as well as the reduced $\chi^{2}$ obtained from the fittings with Gaussian and log-normal models are listed in table \ref{tab:fitting2}.
It is found that the log-normal model is preferred over the Gaussian model.

\begin{deluxetable*}{cccccccc}
	\tablecaption{Best-fitting parameters for the log-normal and Gaussian distribution.\label{tab:fitting2}}
	\tablewidth{0pt}
	\tablehead{
		\colhead{Time bin} & \multicolumn{3}{c}{Log-normal} & \colhead{} & \multicolumn{3}{c}{Gaussian} \\
		\cmidrule(r){2-4}\cmidrule(r){6-8}
		\colhead{} & \colhead{Mean$^{*}$} & \colhead{$\sigma^{*}$} & \colhead{$\chi^{2}_{\rm red}$} & \colhead{} & \colhead{Mean$^{*}$} &  \colhead{$\sigma^{*}$} & \colhead{$\chi^{2}_{\rm red}$}
	}
	\startdata
	{30-d} & $2.28\pm 0.04$ & $0.49\pm 0.04$ & 0.92 & {} & $9.04\pm 0.75$ & $5.27\pm 0.71$ & 1.60 \\
	\enddata
	\tablecomments{
		$^*$: in unit of $\times10^{-8}$ photons cm$^{-2}$\ \rm s$^{-1}$.}
\end{deluxetable*}

\subsection{Searching for Periodicity}
\label{sec:periodicity}

\begin{figure}
	\centering
	\subfigure 
	{\includegraphics[width=13cm]{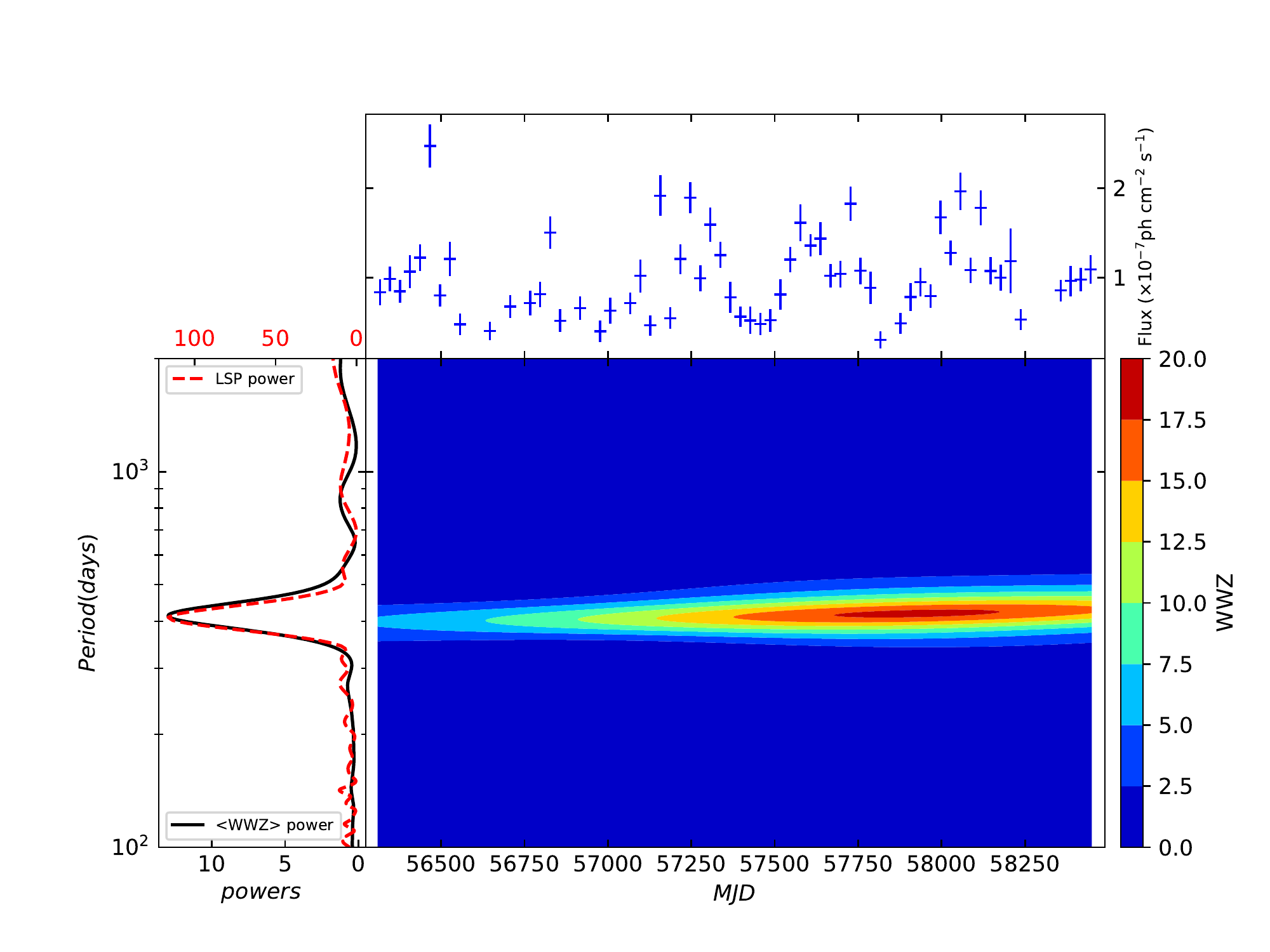}}
	\caption{Upper panel: $\gamma$-ray light curve from MJD 56317 to 58447 with 30-day binning.
	              Right lower panel: two dimensional contour map of the WWZ power spectrum of the $\gamma$-ray light curve.
	              Left lower panel: LSP (red dashed line) and average WWZ (black solid line) powers of the light curve.}\label{fig:wwz}
\end{figure}

\begin{figure}
	\centering
	{\includegraphics[width=11cm]{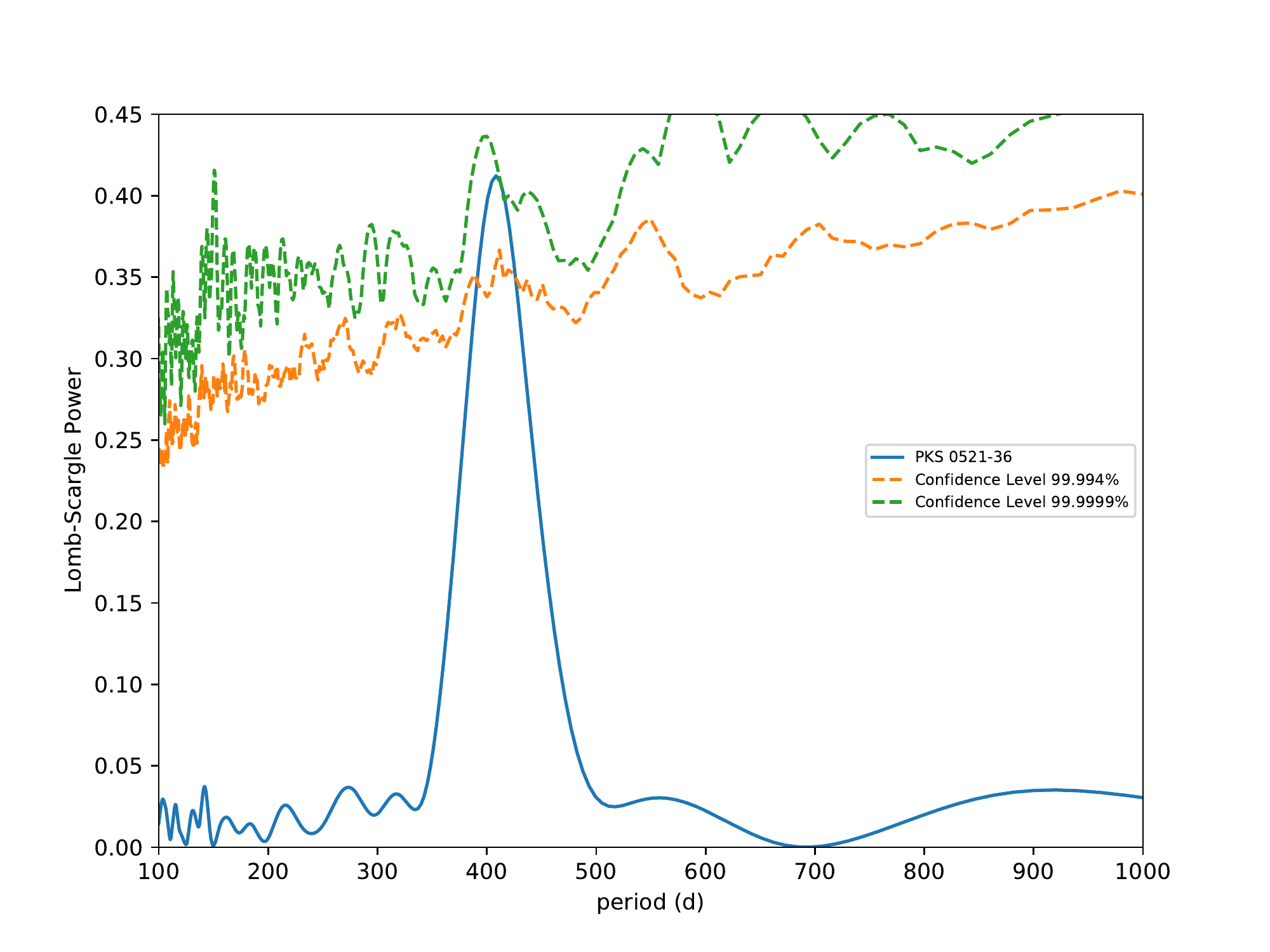}}
	\caption{LSP power calculated from the light curve during MJD 56317-58447 (blue solid line),
	              and the 4$\sigma$ (orange dashed line), 5$\sigma$ (green dashed line) significance curves
	              calculated based on the simulation of $10^{5}$ light curves  with the method given by  \cite{2013MNRAS.433..907E}.}
	              \label{fig:de}
\end{figure}

\begin{figure}
	\centering
	{\includegraphics[width=11cm]{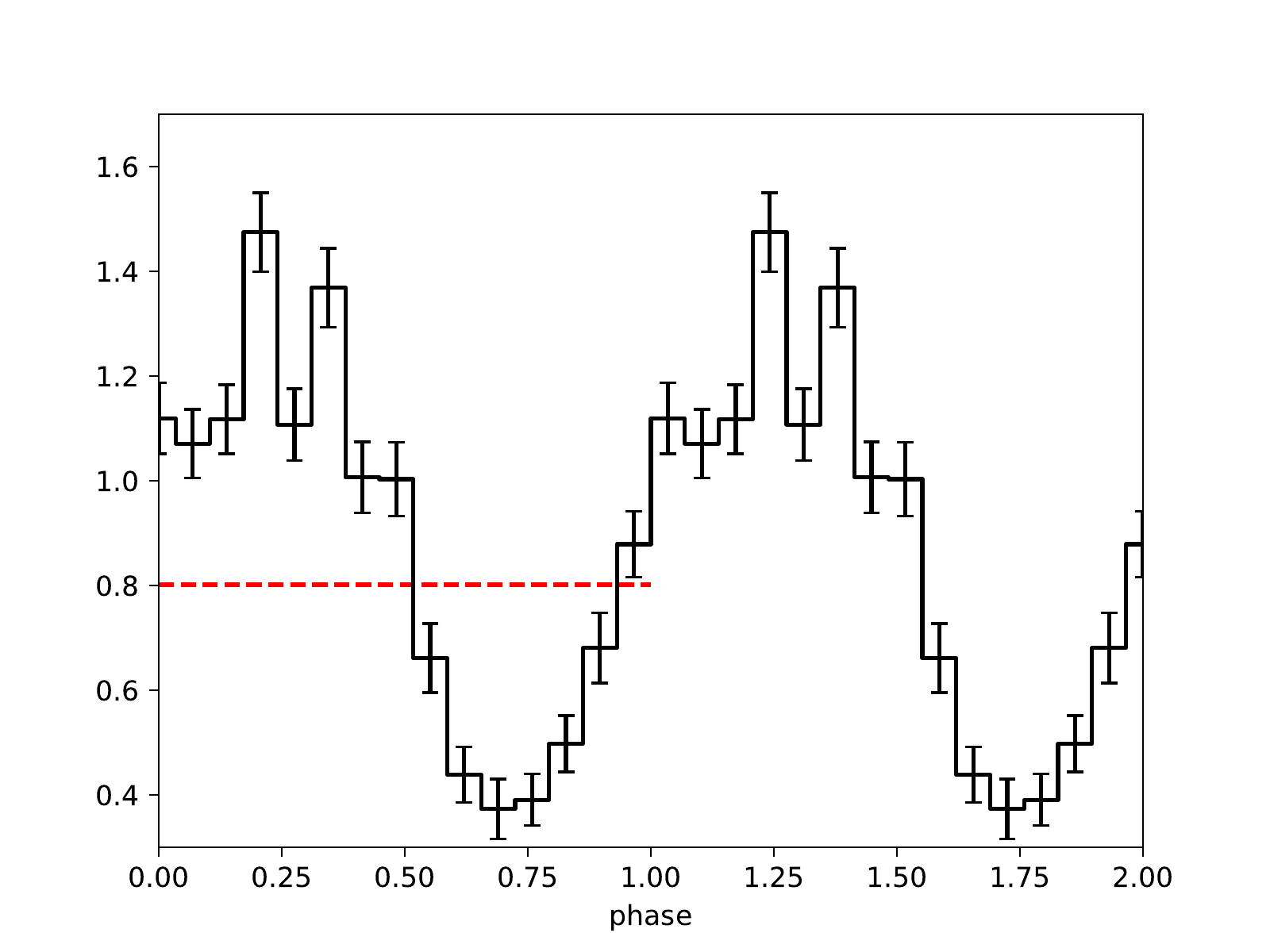}}
	\caption{Folded light curve constructed from binned likelihood
		analysis of the data from MJD 56317-58447 above 100 MeV with a period of 411 days.
		The red dashed line is the constant fitting to the folded light curve in one phase. For clarity, we show two period cycles.}\label{fig:phase}
\end{figure}

\begin{figure}
	\centering
	{\includegraphics[width=11cm]{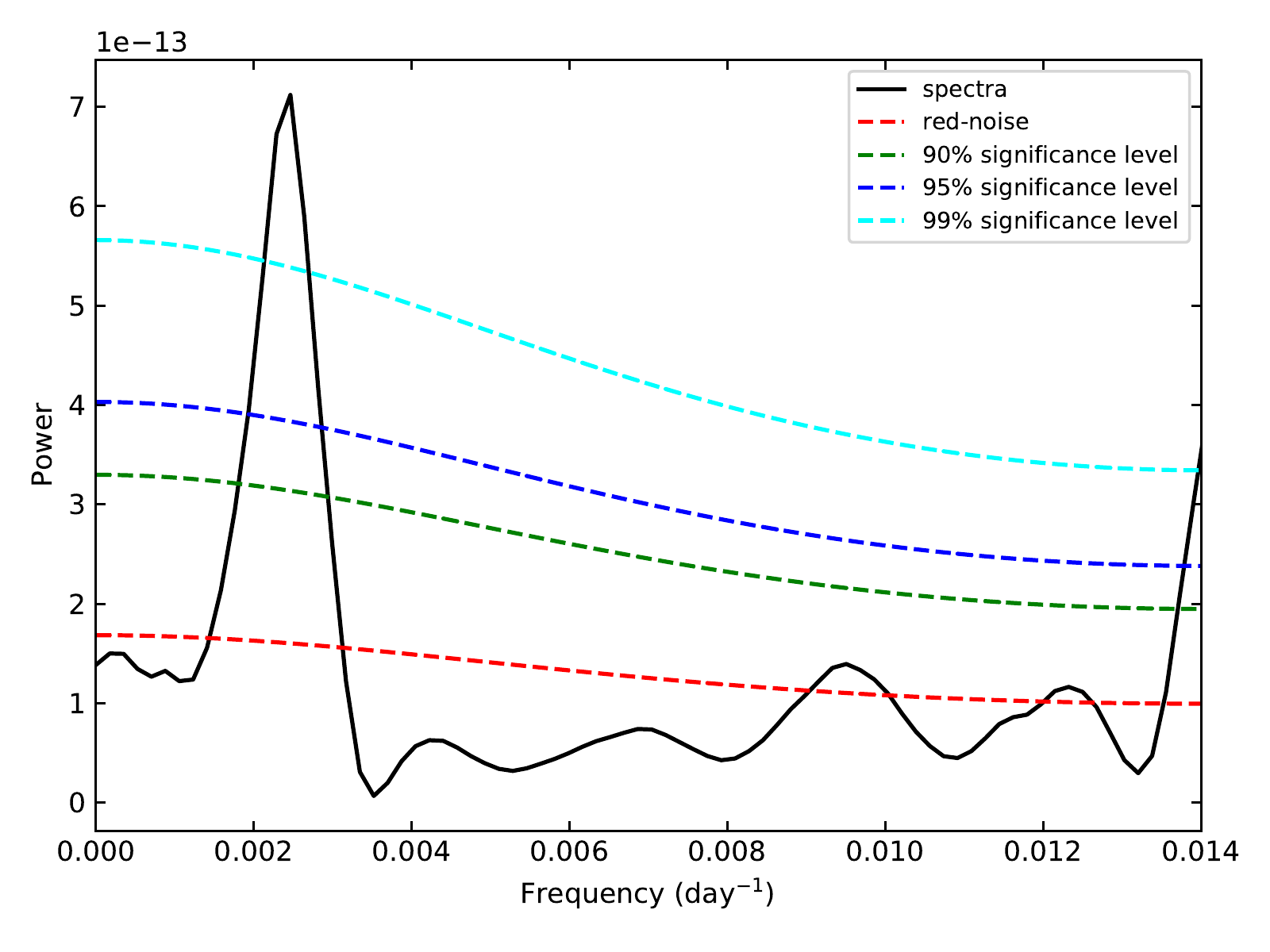}}
	\caption{Results of the periodicity analysis by REDFIT. The black solid line is bias-corrected power
		spectrum. Dashed curves starting from the bottom are the theoretical red-noise spectrum, 90$\%$, 95$\%$ and 99$\%$ significance levels, respectively.
		}\label{fig:REDFIT}
\end{figure}

The Lomb-Scargle periodogram \citep[LSP;][]{1976Ap&SS..39..447L,1982ApJ...263..835S} and the Weighted Wavelet Z-transform \citep[WWZ;][]{1996AJ....112.1709F}
are two of the most employed and best known methods for detection of periodicity in time-series in astronomy.
The LSP and WWZ methods are then applied to the light curve of PKS 0521-36 to search for periodicity.
A clear periodic modulation appears from MJD 56317 to 58447 (Figure~\ref{fig:wwz}). 
A significant peak at $\sim$400 days appears in both LSP and WWZ powers, which indicates a quasi-periodic oscillation (QPO). 
The period is estimated by fitting the power peak with a Gaussian function. It is $411 \pm33$ days given by LSP and $417\pm39$ days given by WWZ.
The bootstrap approximation method is used to calculate the false alarm probability (FAP), 
which is a robust way to
estimate FAP \citep[e.g.,][]{2017MNRAS.471.3036P,2020ApJ...896..134P}.
The FAP is 0.0002 in the LSP power, corresponding to the 99.98\% significance level for the signal.

The algorithm given by \cite{2013MNRAS.433..907E} is also used to evaluate the significance of this QPO signal. 
The properties of observed variation, power spectral density (PSD) and PDF, are well applied in this method. 
The PSD model of the form $P(f)=Af^{-\alpha}$ is used, where $\alpha$ is the power-law index. 
The PDF is in the form of a log-normal distribution. 
The 4$\sigma$ and 5$\sigma$ significance curves are obtained by simulating the 30-day binning light curve $10^{5}$ times, and are shown in Figure \ref{fig:de}. 
The significance of the QPO signal is $\approx5\sigma$.

Further evidence for this QPO is presented in the folded light curve constructed by performing the likelihood analysis method with a 411-day period. 
Phase zero is set at MJD 56317. The variation with the phase is clear (see Figure \ref{fig:phase}).
The fitting to the folded light curve in one phase with a constant results in $\chi^{2}_{\rm red}$=449.26/14,
suggesting a significant variability in the folded light curve.

An additional method, REDFIT \citep{2002CG.....28..421S}, is also employed for the periodicity
detection. This method estimates the red-noise spectrum by fitting the data with a first-order
autoregressive (AR1) process. It can precisely evaluate the significance of the peaks in the PSD against the red-noise background.
We use the program REDFIT3.8e\footnote{\url{https://www.marum.de/Prof.-Dr.-michael-schulz/Michael-Schulz-Software.html}}.
The bias of Fourier transform for unevenly spaced data is also removed in the program.
Two conditions should be satisfied when the program is used for data: 
(1) the data points of light curve should not be too clustered;
and (2) the noise background of light curve can be approximated by an AR1 process. 
Obviously, our data meets the first condition.
Moreover, the result of the non-parametric runs test for our data is rtest = 15 (inside the 98$\%$ acceptance region [9,22]),
which indicates that the noise background spectrum is consistent with an AR1 model \citep{2002CG.....28..421S}.
Therefore, REDFIT can be used to detect periodicity in our data. 
We set $n_{50}$=2 and select a hanning window to reduce spectral leakage \citep{1997CG.....23..929S}.
The results show that there is a peak with more than 99$\%$ significance level, and the corresponding periodicity is 440.9$\pm78.6$ days (Figure \ref{fig:REDFIT}). 
This period is consistent with the LSP and the WWZ results.
Note that the REDFIT method only provides a
maximum of significance of 99\%.

\subsection{Gaussian Process Modeling the $\gamma$-ray Light Curve}
\label{sec:celerite}

Besides LSP, WWZ and REDFIT, an alternative approach for the analysis of astronomical variability is the Gaussian process modeling light curves, which treats the variability in the time rather than frequency domain \citep[e.g.,][]{2009ApJ...698..895K,2014ApJ...788...33K,    
	2018MNRAS.476L..55L}. The PSD can be constructed by using modeling results. One popular Gaussian process model is the continuous time autoregressive moving average (CARMA) model developed by \citet{2014ApJ...788...33K}, which has been applied to $\gamma$-ray variability of blazars \citep[e.g.,][]{2018ApJ...863..175G,2019ApJ...885...12R,2021ApJ...907..105Y}. Here a newly developed Gaussian process model {\it celerite} \citep{2017AJ....154..220F}, is used to investigate the $\gamma$-ray variability of PKS 0521-36.

In this work, a stochastically driven damped simple harmonic oscillator (SHO) is used, and its differential equation is given by 
\begin{equation}\label{eq5}
\left[\frac{d^{2}}{dt^{2}}+\frac{\omega_{0}}{Q}\frac{d}{dt}+\omega_{0}^{2}\right]y(t)=\epsilon(t)\;,
\end{equation}
where $\omega_{0}, Q, \epsilon(t)$ are the frequency of the undamped oscillator, the quality factor of the oscillator, and a stochastic driving force (white noise here), respectively. The PSD in the case is
\begin{equation}\label{eq6}
S(\omega)=\sqrt{\frac{2}{\pi}}\frac{S_{0}\omega_{0}^{4}}{(\omega^{2}-\omega_{0}^{2})^{2}+\omega^{2}\omega_{0}^{2}/Q^{2}}\;,
\end{equation}
where $S_{0}$ is proportional to the power when $\omega=\omega_{0}$.

A specific model for a time series can be built up with a
mixture of determined number of SHO terms.  
The Markov Chain Monte Carlo (MCMC) algorithm \citep{2013PASP..125..306F} is used to 
perform fitting to light curve. From MCMC sampling, we can evaluate values and uncertainties of model parameters.
 In detail, $5\times10^{4}$ samples are generated in our analysis, and the first $2\times10^{4}$ samples are taken as burn-in sampling.

\begin{figure}
	\centering
	{\includegraphics[width=11cm]{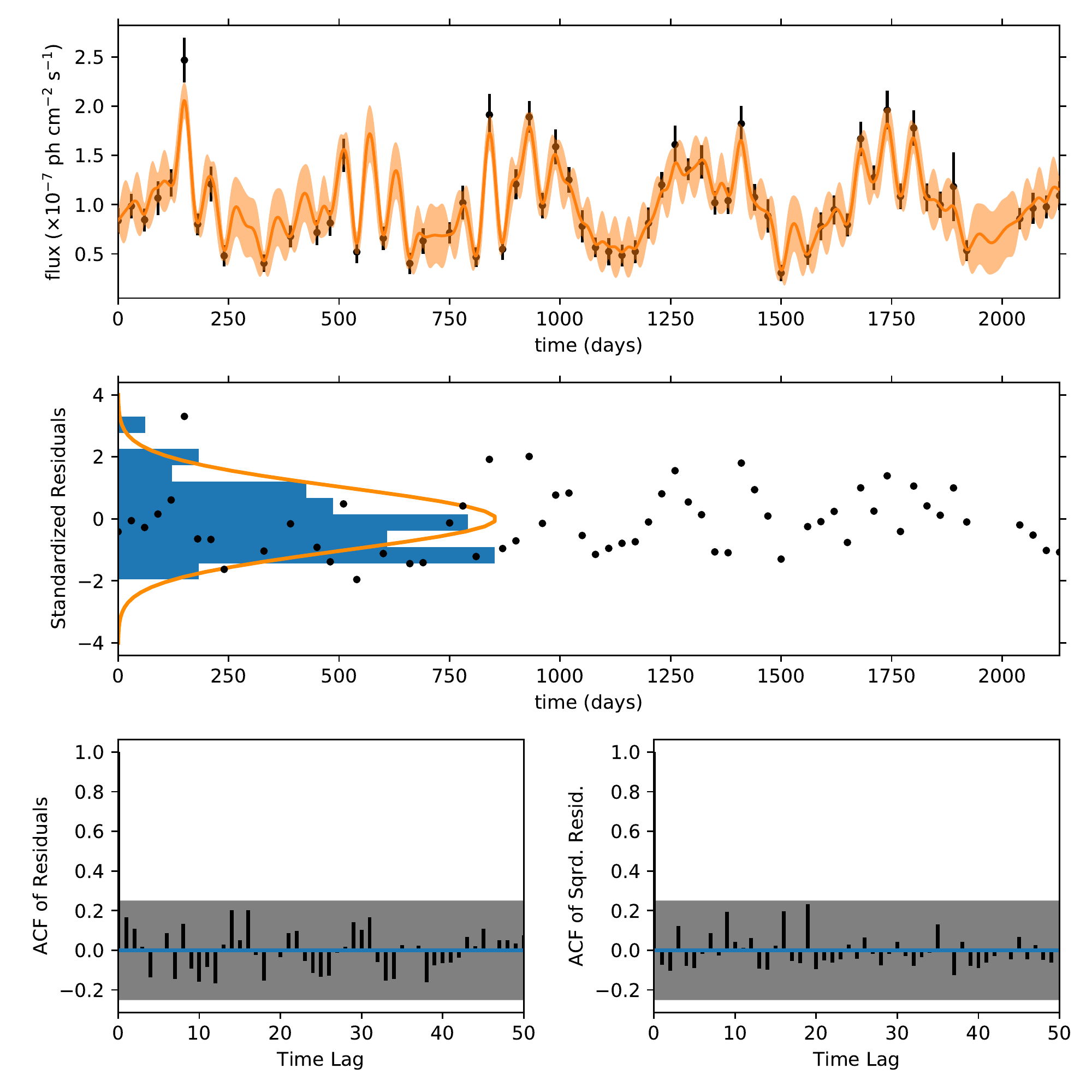}}
	\caption{{\it celerite} modeling results for the 30-day binning $\gamma$-ray light curve of PKS 0521-36 from MJD 56317 to 58447; 
	                      in the top plot, black points with errors are LAT data, and the bright orange line is the best fit to the data with the {\it celerite} model;
	                               black dots, blue histogram and orange line in the middle plot are standardized residuals of each bin, scaled standardized residuals distribution,
	                               and the expected normal distribution (scaled), respectively; the ACFs of residuals and squared residuals are given in the bottom plot where the
	                               gray region is the 95$\%$ confidence limit of the white noise. 
	}\label{fig:celerte30}  
\end{figure}

\begin{figure}
	\centering
	 {\includegraphics[width=14cm]{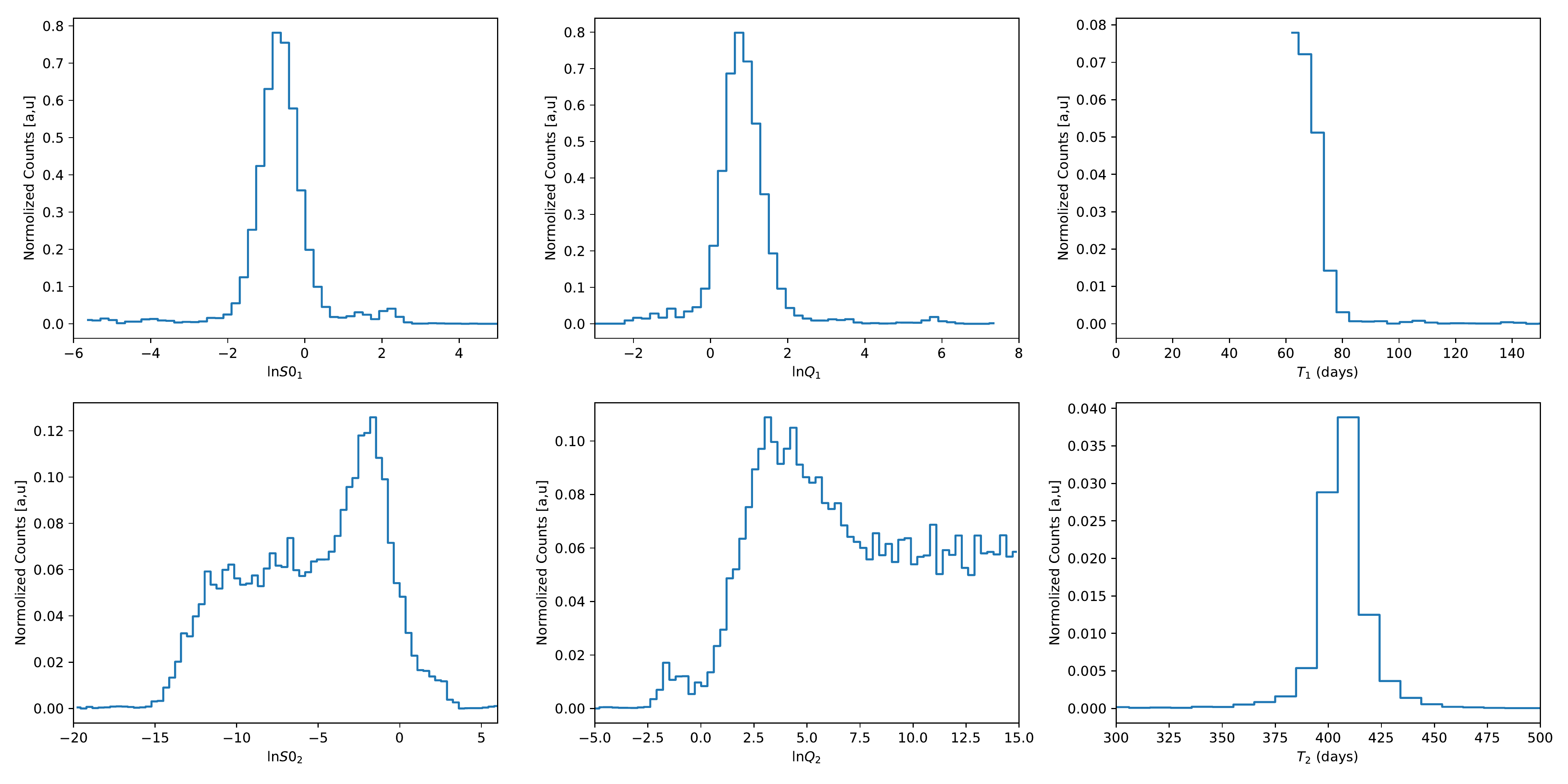}}
	\caption{Posterior probability densities of parameters obtained from {\it celerite} modeling results.
	}\label{fig:celerte_param}  
\end{figure}

\begin{figure}
	\centering
	{\includegraphics[width=11cm]{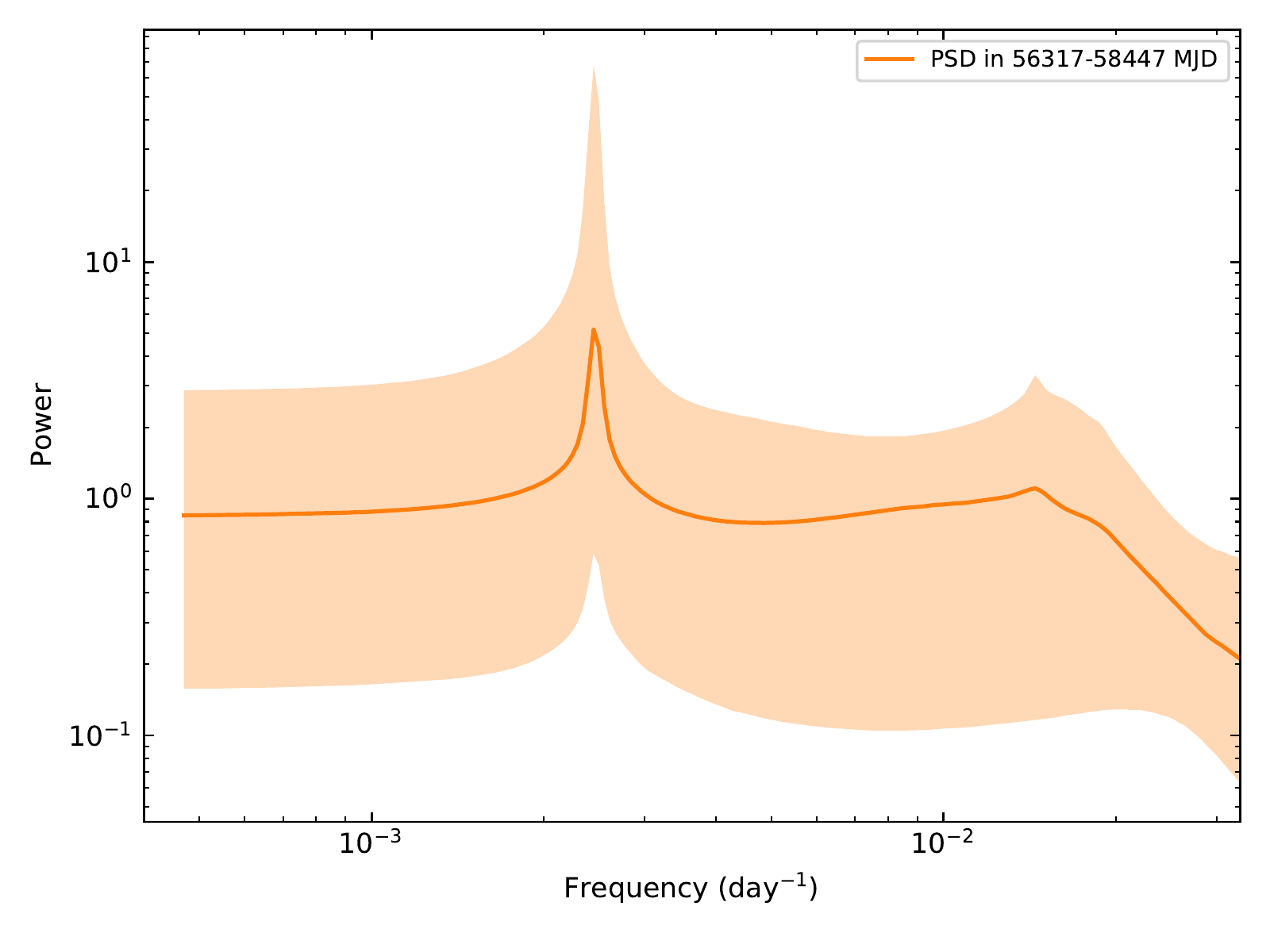}}
	\caption{PSD constructed from the {\it celerite} modeling results for the light curve during MJD 56317-58447. Bright color region is the 68$\%$ confidence interval. 
	A distinct peak appears at $\sim400$ days.}
	\label{fig:psd}  
\end{figure}

Figure~\ref{fig:celerte30} shows the fitting results to the 30-day binning light curve of PKS 0521-36 from MJD 56317 to 58447 with the {\it celerite} model (SHO$\times 2$). 
For the modeling results, we use Kolmogorov-Smirnov (KS) test to check the consistency between the standardized residuals 
and a normal distribution with mean zero and standard deviation of unity. 
The p-values of the test is p=0.23, which means this consistency is credible \citep{2021ApJ...907..105Y}.
The auto-correlation functions (ACFs) of the residuals and squared residuals in Figure~\ref{fig:celerte30} are almost inside the 95$\%$ confidence limits,
so that the correlation structures have been captured by the model.
These results indicate that the $\gamma$-ray variability can be described well by the two-SHO model.
The posterior probability densities for the model parameters are shown in Figure~\ref{fig:celerte_param} and the value of these parameters are given in Table~\ref{tab:celerite_paramfitting}.
A QPO feature with period of $\sim400$ days appears in the PSD constructed from the modeling results (Figure~\ref{fig:psd}).

\begin{deluxetable*}{ccccc}
	\tablecaption{Parameters for the {\it celerite} modeling results corresponding to Figure~\ref{fig:celerte_param}.\label{tab:celerite_paramfitting}}
	\tablewidth{0pt}
	\setlength{\tabcolsep}{8mm}{
	\tablehead{
		\colhead{Time} & \colhead{SHO No.} & \colhead{$lnS_{0}$} & \colhead{$lnQ$} & \colhead{$T$} \\
		\colhead{(MJD)} & \colhead{} & \colhead{} & \colhead{} & \colhead{(days)}
	}
	\decimalcolnumbers
	\startdata
	56317-58447 & 1 & $-0.65_{-0.53}^{+0.54}$ & $ 0.80_{-0.5}^{+0.55}$ & $66.5_{/}^{+5.18}$  \\
	& 2 & $ -4.78_{-5.64}^{+3.57}$ & $6.60_{-3.77}^{+5.64}$ & $405.93_{-10.32}^{+9.68}$
	\enddata
	\tablecomments{(1) time period, (2) number of SHO terms, (3)-(5) parameters in Equations \ref{eq6}, T=$2\pi/\omega_{0}$. }}
\end{deluxetable*}

There is an aperiodic component and a periodic component in the $\gamma$-ray variability during the time range of MJD 56317-58447.
The period of the QPO is estimated as $406\pm10$ days which is consistent with the results given by LSP, WWZ and REDFIT.
The corresponding quality factor of the oscillator is large (Table~\ref{tab:celerite_paramfitting}), suggesting that the QPO is significant \citep{2017AJ....154..220F}.

\subsection{Short Timescale Variability}
\label{sec:variability}

\begin{figure}
	\centering
	{\includegraphics[width=11cm]{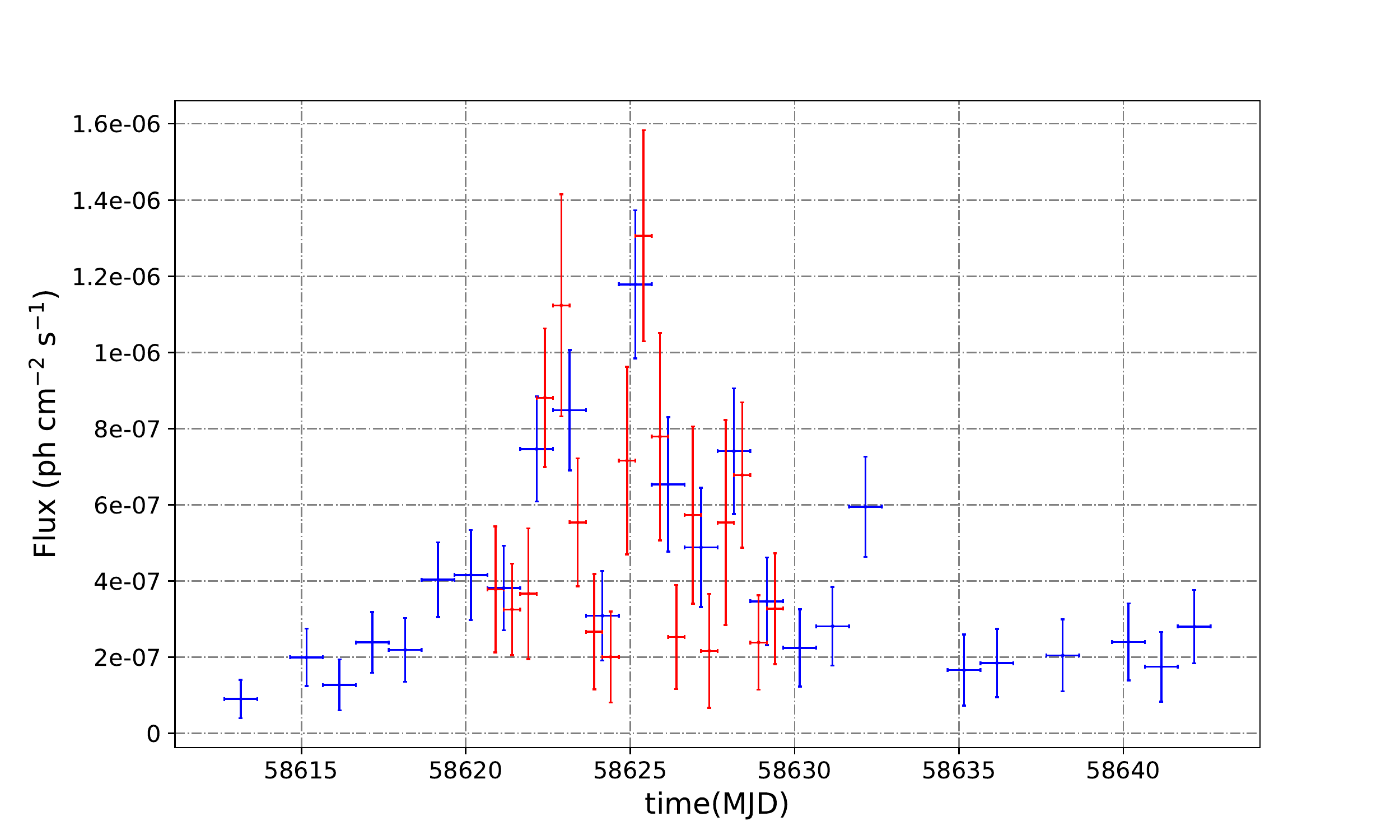}}
	\caption{One-day binning light curve (blue points) and 12-hour binning light curve (red points) for the bright flare during 2019 May 9-June 8 (MJD 58612-58642).} 
	\label{fig:shortcurve}  
\end{figure}

\begin{figure}
	\centering
	{\includegraphics[width=11cm]{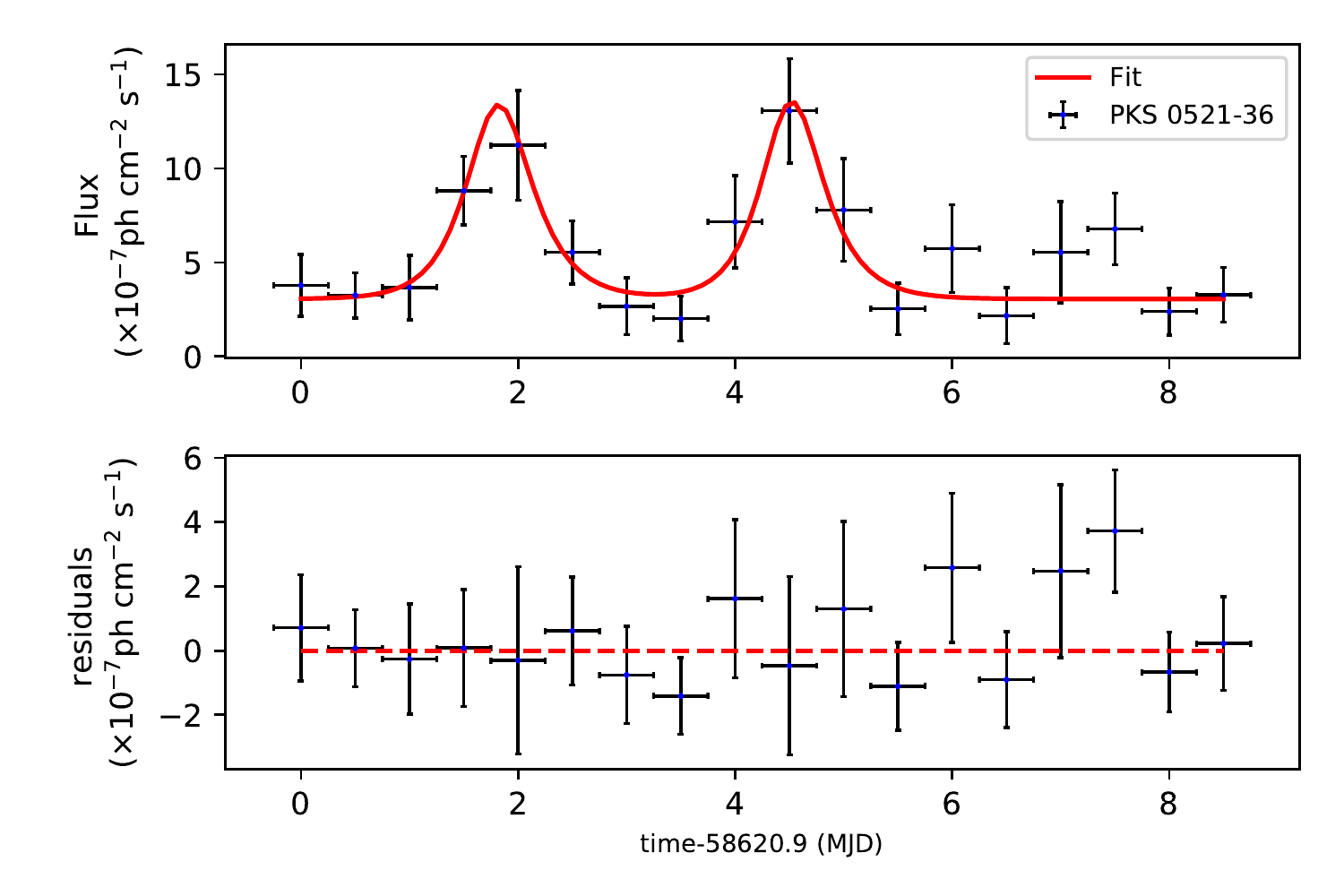}}
	\caption{Best-fitting results for the 12-hour binning light curve showed in Figure~\ref{fig:shortcurve} with Equation~\ref{eq4},
	             and the distribution of the residuals obtained in the fitting (bottom panel) which is fitted by a constant (red dashed line).} 
	\label{fig:lightcurve3}  
\end{figure}

Intraday $\gamma$-ray variability of PKS 0521-36 has been found in the LAT data \citep{2015MNRAS.450.3975D,2019A&A...627A.148A}.
A recent bright $\gamma$-ray flare occurred in 2019 May 9-June 8 (MJD 58612-58642) has not been investigated.
In our work, intraday variability during this flare is observed in the one-day and half-day binning light curves (Figure~\ref{fig:shortcurve}).
Here we use unbinned maximum likelihood method to construct the light curves, 
and select time bins having TS$\geqslant$ 4.
This flare is not bright enough to use the selection criteria of TS$\geqslant$ 25 for such a short time bin. 

In order to determine variability timescale, the 12-hour binning light curve is fitted by using the following function \citep{2010ApJ...722..520A}
\begin{equation}\label{eq4}
F(t)=F_{\rm c}+F_{0}\left( e^{\frac {t_{0}-t}{T_{\rm r}}}+e^{\frac {t-t_{0}}{T_{\rm d}}}\right)^{-1}\;,
\end{equation}
where $F_{\rm c}, F_{0}, t_{0}, T_{\rm r}$ and $T_{\rm d}$ are a constant flux, peak flux, peak time, and rise and decay times, respectively. 
This function can be used to describe one-peak flare. The light curve with multi-peaks is considered as a whole, and is fitted by using the model comprising several components of Equation~\ref{eq4}. The best-fitting results are presented in Figure \ref{fig:lightcurve3},
and the parameters are given in Table \ref{tab:fitting1}. The residuals are fitted by the constant of zero successfully with $\chi^{2}_{\rm red}=0.58$, suggesting that our fitting to the light curve is acceptable. It is found $T_{\rm r}\sim T_{\rm d}\sim$ 7 hours. This is the third short timescale $\gamma$-ray flare in PKS 0521-36.
However, the loose data selection criteria of TS$\geqslant$4 decreases the significance of the derived timescale. 

The SEDs during 2019 May 9 - June 8 (MJD 58612-58642) and 2012 October 12 - November 11 (MJD 56212-56242) are also produced. 
The spectral parameters are listed in Table~\ref{tab:fitting}, together with the parameters for the average SED shown in Figure~\ref{fig:sed}. It is found that during the two flares, the SEDs are the same, which can be described by a power-law form, different from the average LP SED.

\begin{deluxetable*}{cccccccc}
	\tablecaption{Best-fitting parameters for the 12-hour light curve in Figure~\ref{fig:lightcurve3}.\label{tab:fitting1}}
	\tablewidth{0pt}
	\tablehead{
		\colhead{Time} & \colhead{Peak No.} & \colhead{$F_{\rm c}$} & \colhead{$F_{0}$} & \colhead{$t_{0}$} & \colhead{$T_{\rm r}$} & \colhead{$T_{\rm d}$} & \colhead{$\chi^{2}_{\rm red}$} \\
		\colhead{(MJD)} & \colhead{} & \colhead{} & \colhead{} & \colhead{(days)} &  \colhead{(days)} & \colhead{(days)} & \colhead{}
	}
\decimalcolnumbers
	\startdata
	58620.9-58629.4 & p1 & $3.06\pm 0.54$ & $ 20.63\pm 8.81$ & $1.80\pm0.31$ & $0.25\pm0.27$ &  $0.29\pm0.18$  & 0.90 \\
	 & p2 & & $ 20.95\pm 5.55$ & $4.5^{*}$ & $0.24\pm0.11$ &  $0.28\pm0.12$ &
	\enddata
	\tablecomments{(1) time period, (2) peak number, (3)-(7) parameters in Equations \ref{eq4} , (8) the reduced $\chi^2$, (3)(4) in unit of $10^{-7}$ photons cm$^{-2}$ s$^{-1}$.
	  The value with $*$ represents that it is fixed in fitting. }
\end{deluxetable*}

\begin{deluxetable*}{cccccc}
	\tablecaption{Best-fitting spectral parameters in four periods.\label{tab:fitting}}
	\tablewidth{0pt}
	\tablehead{
		\colhead{Time period} & \colhead{Time period} & \colhead{$\alpha$} & \colhead{$\beta$} & \colhead{TS} & \colhead{Flux} \\
		\colhead{(MJD)} & \colhead{(UT)} & \colhead{} & \colhead{} & \colhead{} & \colhead{($10^{-7}$ photons cm$^{-2}$s$^{-1}$)}
	}
	\startdata
	54682-59287 & 2008-08-04/2021-03-29 & $2.43\pm0.02$ & $0.05\pm 0.01$ & 18447.2 & $1.06\pm 0.02$ \\
	56212-56242 & 2012-10-12/2012-11-11 & $2.41\pm0.05$ & $--$ & 1363.98 & $3.70\pm 0.21$ \\
	58612-58642 & 2019-05-09/2019-06-08 & $2.39\pm0.08$ & $--$ & 851.68 & $3.22\pm 0.24$ \\
	56317-58447 & 2013-01-25/2018-11-25 & $2.46\pm0.03$ & $0.06\pm 0.02$ & 6188.17 & $0.91\pm 0.02$ \\
	\enddata
\end{deluxetable*}

\section{Discussion} \label{sec:discussion}

Our results show that the PDF obtained from the $\gamma$-ray light curve of PKS 0521-36 from 2008 August 4 to 2021 March 29 is preferentially log-normally distributed. The log-normal distribution of flux is found in X-ray binaries \citep[e.g.,][]{2001MNRAS.323L..26U}, X-ray NLSy1s \citep[e.g.,][]{2005MNRAS.359..345U}, and blazars \citep[e.g.,][]{2016A&A...591A..83S,2018RAA....18..141S,2020A&A...634A.120A,2020ApJ...891..120B}.
Similar as X-ray binaries, the log-normal distribution of $\gamma$-ray fluxes may indicate an influence of the accretion disk on the jet \citep{2009A&A...503..797G}, i.e., the multiplicative behavior in an accretion disc impacting the jet. In this scenario, the variability timescale is dominated by the viscous timescale of the accretion disc \citep{2019Galax...7...28R}. 
However, the tentative timescale of $\sim7$ hours $\gamma$-ray variability found in PKS 0521-036 is unlikely achieved in this scenario \citep{2006ASPC..360..265C}.
The cascade-related emission processes may also lead to a log-normal distribution of fluxes \citep{2019Galax...7...28R}, for instance, the cascade emission in lepto-hadronic jet model \citep[e.g.,][]{1993A&A...269...67M,2013ApJ...768...54B,2015MNRAS.447.2810Y}. In this case, the observed GeV emission is produced by the cascade processes, which is feasible. However, it would require an extremely high jet power \citep[e.g.,][]{2015MNRAS.448.2412P,2020Galax...8...72C}.

A simple and natural scenario is that the log-normal distribution of variability results from the emitting particles with the log-normal distribution \citep{2018MNRAS.480L.116S,2019Galax...7...28R}. 
This interpretation is associated with the fluctuations in the particle acceleration rate \citep{2018MNRAS.480L.116S}. 
The average LP SED suggests an average LP electron distribution. 
Both the log-normal and LP electron distributions could be the result of stochastic acceleration \citep{2011ApJ...739...66T}. 

Possible $\gamma$-ray QPOs have been reported in blazars \citep[e.g.,][]{2015ApJ...810...14A,2017MNRAS.471.3036P,2020ApJ...896..134P,2014ApJ...793L...1S,2016AJ....151...54S,2017A&A...600A.132S,2017ApJ...835..260Z,2017ApJ...845...82Z,2018NatCo...9.4599Z},
although cautions have been raised on the significance of these QPOs \citep[e.g.,][]{2019MNRAS.482.1270C,2020ApJ...895..122C,2020A&A...634A.120A,2021ApJ...907..105Y}.
It is well known that blazar jets are highly beamed.
So far, no $\gamma$-ray QPO is reported in non-blazar AGNs.

Here, we report a $\gamma$-ray QPO with a significance of $\sim 5\sigma$ in non-blazar AGN PKS 0521-36. 
The period is $\sim$1.1 years. This QPO behavior only appears in the temporal range from MJD 56317 to 58447 which is between two outbursts (occurred in 2012 October and 2019 May respectively). 
This significant QPO is obtained by three different periodicity detection methods (LSP, WWZ and REDFIT).
Evidence for this QPO is also found in the Gaussian process modeling the 30-day binning $\gamma$-ray light curve. 
It is confirmed that the jet of PKS 0521-36 is a mildly beamed \citep{2016A&A...586A..70L,2019A&A...627A.148A}.
Therefore, the QPO of PKS 0521-36 is the first $\gamma$-ray QPO found in a mildly beamed jet.
The average SED during MJD 56317-58447 is described by a LP model (see Table~\ref{tab:fitting}) which is different from the power-law SEDs in the two outbursts. 

The detection of a helicoidal motion in the optical jet of PKS 0521-36 was reported \citep{2017MNRAS.470L.107J}, 
which may be associated with helical magnetic field or jet precession. 
A simultaneous VLBI core flux activity during the $\gamma$-ray flare in 2010-2011 was reported by \citet{2019A&A...627A.148A}. 
This suggests that the $\gamma$-ray emission region is located inside the radio core rather in the jet. 
$\gamma$-ray variability on timescale of $\sim$7 hours is observed before and after the QPO.
If the $\gamma$-ray photons produced in a single region, the helicoidal motion along the jet at pc scales \citep{2017MNRAS.470L.107J} could not lead to the $\gamma$-ray QPO of PKS 0521-36.

It seems that the outburst during 2012 October plays a key role in producing this QPO.
It is not clear whether  the outburst during 2019 May destroys this QPO or triggers another QPO.
The data collected in future several years will answer this question.

\section{Summary} \label{sec:summary}

We have carried out a temporal analysis of the $\gamma$-ray emissions from an interesting non-blazar 
AGN PKS 0521-36 with the {\it Fermi}-LAT data from 2008 August 4 to 2021 March 29. Our main results are as follows. 

$(\romannumeral1)$ The PDF of the $\gamma$-ray fluxes is more consistent with a log-normal distribution rather than a Gaussian distribution.
It may be due to the log-normal distribution of emitting electrons, which could be the result of stochastic acceleration. 
Moreover, the total average LP SED also can be produced by the stochastic acceleration.

$(\romannumeral2)$ Variability on timescale of $\sim7$ hours is found during the outburst in 2019 May. 
It is the third short timescale $\gamma$-ray flare in PKS 0521-36. 
Such a variability indicates that the $\gamma$-ray emission region is compact.
However, it should be noted that the timescale may be in low significance.

$(\romannumeral3)$ A QPO with the period of $\sim$1.1 years is found at $\sim5\sigma$ confidence level in the time interval (MJD 56317-58447) between two outbursts. 
Four different methods are used to examine the robustness of the QPO.
All the methods give significant evidence for this QPO. 
This is the first $\gamma$-ray QPO found in a mildly beamed jet.
The SED during the QPO interval is different from that during the two $\gamma$-ray outbursts in 2012 October and 2019 May.
It is argued that the two $\gamma$-ray outbursts play a key role in the formation of this $\gamma$-ray QPO.
The $\gamma$-ray variability during the QPO interval can be well modeled by two SHO components,  an aperiodic component and a periodic component.

\acknowledgments
We thank the anonymous reviewer for constructive suggestions. We acknowledge financial support from National Key R\&D Program of China under grant No. 2018YFA0404204,
 the National Science Foundation of China 11803081,
 and the joint foundation of Department of Science and Technology of Yunnan Province and Yunnan University [2018FY001(-003)].
The work of D. H. Yan is also supported by the CAS Youth
Innovation Promotion Association and Basic research Program of Yunnan Province (202001AW070013).

{\it Facility:} Fermi(LAT)

{\it Software:} Fermitools-conda, celerite \citep{2017AJ....154..220F}, emcee \citep{2013PASP..125..306F}, REDFIT \citep{2002CG.....28..421S}, NumPy \citep{2020NumPy-Array}, Matplotlib \citep{2007CSE.....9...90H}, Astropy \citep{2013A&A...558A..33A,2018AJ....156..123A}, SciPy \citep{2020SciPy-NMeth}, DELCgen \citep{2016ascl.soft02012C}.

\bibliography{PKS0521}{}
\bibliographystyle{aasjournal}

\end{document}